# Manipulation of magnetic solitons under the influence of DMI gradients

Rayan Moukhader[1,2], Davi Rodrigues[3], Eleonora Raimondo[1], Vito Puliafito[3], Bruno Azzerboni[4], Mario Carpentieri[3], Abbass Hamadeh[2,5], Giovanni Finocchio[1,6*], Riccardo Tomasello[3*]

[1]Department of Mathematical and Computer Sciences, Physical Sciences and Earth Sciences, University of Messina, I-98166, Messina, Italy

[2]Multi-Disciplinary Physics Laboratory, Faculty of Sciences, Lebanese University, Beirut, Lebanon

[3]Department of Electrical and Information Engineering, Politecnico of Bari, 70125 Bari, Italy

[4]Department of Engineering, University of Messina, 98158 Messina, Italy

[5]Fachbereich Physik and Landesforschungszentrum OPTIMAS, Technische Universitat Kaiserslautern, 67663 Kaiserslautern, Germany

[6]Istituto Nazionale di Geofisica e Vulcanologia (INGV), Via Vigna Murata 605, 00143 Roma, Italy

**Abstract**

Magnetic solitons are promising for applications due to their intrinsic properties such as small size, topological stability, ultralow power manipulation and potentially ultrafast operations. To date, research has focused on the manipulation of skyrmions, domain walls, and vortices by applied currents. The discovery of new methods to control magnetic parameters, such as the interfacial Dzyaloshinskii-Moriya interaction (DMI) by strain, geometry design, temperature gradients, and applied voltages promises new avenues for energetically efficient manipulation of magnetic structures. The latter has shown significant progress in 2d material-based technology. In this work, we present a comprehensive study using numerical and analytical methods of the stability and motion of different magnetic textures under the influence of DMI gradients. Our results show that under the influence of linear DMI gradients, Néel and Bloch-type skyrmions and radial vortex exhibit motion with finite skyrmion Hall angle, while the circular vortex undergoes expulsion dynamics. This work provides a deeper and crucial understanding of the stability and gradient-driven dynamics of magnetic solitons, and paves the way for the design of alternative low-power sources of magnetization manipulation in the emerging field of 2d materials.

## 1. Introduction

Over the last decades, information technology has become eminently relevant for our everyday lives. The emergent discipline of spintronics together with mathematical concepts of topology have introduced novel paradigms based on exotic non-uniform magnetic configurations at the nanoscale level, opening new horizons for designing the next generation of ultralow power devices in IT applications. The competition of interfacial effects, such as Dzyaloshinskii-Moriya interaction (DMI)[1,2] and perpendicular anisotropy[3], play the main role in stabilizing and influencing the dynamics of these magnetic solitons. The most prominent example of these textures is the magnetic skyrmion[4] which has attracted much attention in information storage and logic technologies since its first experimental observation of a Bloch skyrmion in the B20 compounds[5–7] driven by bulk DMI (b-DMI). Later, skyrmions have been observed in a variety of thin film materials with a perpendicular easy-axis of the magnetization and interfacial DMI (i-DMI), including Heavy Metal (HM)/single-layer ferromagnet (FM)/oxide[8,9], HM1/FM/HM2 multilayers[10,11], HM/ferrimagnet/oxide multilayers[12,13] combinations of FM/ferrimagnets[14,15], as well as synthetic antiferromagnets (SAFs)[16–18]. A pioneering application of skyrmions was the racetrack memory[19,20], however many others have been proposed, such as oscillators[21–24], detectors[25], random bit generator[26], logic gates[27], and recently unconventional applications in probabilistic computing[28,29], as well as in neuromorphic computing[30]. Additionally, materials exhibiting in-plane easy-axis are able to host a different type of soliton, i.e. vortex[31], whose voltage and field-driven manipulation have demonstrated applications to memory and sensors devices[32,33]. Magnetic vortices are a typical ground state of ferromagnetic dots and are characterized by two properties: polarity and chirality. The polarity refers to the direction of the vortex core being either upward (p=+1) or downward (p=-1), and the chirality refers to the configuration of the magnetization around the vortex core, which can be circular (clockwise or anti-clockwise) or radial (inwards or outwards)[34,35], the latter promoted by the i-DMI. A wide variety of approaches were proposed for the manipulation of skyrmions and vortices, including electrical currents through the conventional spin-orbit torques (SOT)[36,37], external field gradients[38,39], anisotropy gradients[40,41] as well as thermal gradients[42–44]. Notably, parameters gradients prove to be very useful in inducing dynamics in the case of electrical insulators compared to electrical manipulation. However, the effect of linear gradients of parameters has been deeply investigated for skyrmions[44,45], while less attention has been dedicated to vortices. Hence, we aim to extend this alternative method to the manipulation of other magnetic solitons, such as circular and radial vortices, in different materials and geometries. Recently, 2d materials have been hailed as building blocks for the next generation electronic devices[46]. Several 2d materials, such as CrI3[47], Fe3GeTe2[48], and VSe2[49], have shown potentials for the stabilization of non-collinear magnetic textures and spintronic applications. Interesting features of 2d materials, in particular in CrI3 and in Janus monolayers Cr(I,X)3 (X = Br, Cl), is the voltage and strain control of magnetic parameters relevant for magnetic textures stability[50–52]. Here, we present a theoretical analysis carried out via systematic micromagnetic simulations and corroborated by the Thiele's formalism. We consider different magnetic solitons (Néel and Bloch skyrmions, as well as circular and radial vortices) in different geometries (circular and rectangular samples) and two materials (thin film CoFeB [34] and 2d CrI3 [53]). We demonstrate the driven-motion of the considered solitons via a linear DMI gradient, which, in principle, can be controlled by different methods, such as voltage-gating [50–52]. We notice that CrI3 is well-known to exhibit an antiferromagnetic interlayer coupling, in a bilayer geometry, that can be modeled as a standard Ruderman–Kittel–Kasuya–Yosida (RKKY) interaction[54]. Therefore, we considered the DMI-gradient-driven dynamics also in SAFs. We first perform static simulations to obtain the stability range of the solitons as a function of the DMI. We show that the typical values of DMI to stabilize textures in 2d materials can be two orders of magnitude smaller than for standard thin-films. Second, we study the DMI-gradient driven motion of the different magnetic solitons. We observe that both skyrmions and radial vortices move under the DMI gradient, with a damping-dependent trajectory. We also show how the i-DMI is able to drive circular vortices expulsion. This study provides a comprehensive description of the behavior of each stable topological texture in the different materials and geometries in the presence of DMI gradients. Our results can stimulate further experimental activities in the manipulation of solitons by material gradients for low-power spintronic devices.

## 2. Modelling and Samples Properties

*2.1 Micromagnetic modelling.* The micromagnetic computations were performed with a state-of-the-art micromagnetic solver, PETASPIN, a local CUDA-native and multi-GPU solver benchmarked against the standard problems of micromagnetic community[55]. Our code uses the Adams-Bashforth[56,57] method to numerically integrate the Landau-Lifshitz-Gilbert (LLG) equation,

$$\frac{d\mathbf{m}}{d\tau} = -(\mathbf{m} \times \mathbf{h}_{\text{eff}}) + \alpha_G \left( \mathbf{m} \times \frac{d\mathbf{m}}{d\tau} \right) \quad (1)$$

where $\alpha_G$ is the Gilbert damping, $\mathbf{m} = \mathbf{M}/M_s$ is the normalized magnetization vector, and $\tau = \gamma_0 M_s t$ is the dimensionless time, with $\gamma_0$ being the gyromagnetic ratio, and $M_s$ the saturation magnetization. The effective field $\mathbf{h}_{\text{eff}}$ includes the exchange (A being the exchange constant), DMI (D being the DMI constant), uniaxial anisotropy ($K_u$ being the anisotropy constant) and magnetostatic field.

The b-DMI energetic density expression is:

$$\varepsilon_{b-DMI} = D\mathbf{m} \cdot \nabla \times \mathbf{m} \tag{2}$$

and, making the functional derivative of Eq. (2), the b-DMI effective field is derived:

$$\mathbf{h}_{b-DMI} = -\frac{1}{\mu_0 M_S}\frac{\delta \varepsilon_{b-DMI}}{\partial \mathbf{m}} = -\frac{2D}{\mu_0 M_S}\nabla \times \mathbf{m}. \tag{3}$$

The boundary conditions related to the b-DMI are expressed by:

$$\frac{d\mathbf{m}}{dn} = \frac{\mathbf{m} \times \mathbf{n}}{\xi}, \tag{4}$$

where **n** is the unit vector normal to the edge and $\xi = \frac{2A}{D}$ is a characteristic length in presence of the DMI. Analogously, the i-DMI energetic density, effective field expression and boundary conditions are:

$$\varepsilon_{i-DMI} = D\left[m_z \nabla \cdot \mathbf{m} - (\mathbf{m} \cdot \nabla)m_z\right] \tag{5a}$$

$$\mathbf{h}_{i-DMI} = -\frac{2D}{\mu_0 M_S}\left[(\nabla \cdot \mathbf{m})\hat{z} - \nabla m_z\right] \tag{5b}$$

$$\frac{d\mathbf{m}}{dn} = \frac{1}{\xi}(\hat{z} \times \mathbf{n}) \times \mathbf{m} \tag{5c}$$

Where $m_z$ is the z-component of the normalized magnetization, **n** is the unit vector normal to the edge, and the ultra-thin film hypothesis ($\frac{\partial \mathbf{m}}{\partial z} = 0$) is considered [20,57,58]. For skyrmions, we consider both types of DMI, leading to stabilization of Bloch and Néel types. For vortices, we only consider the effect of i-DMI.

It has been experimentally shown that while an odd number of layers CrI3 reveals ferromagnetism, for an even number of layers an antiferromagnetic interlayer coupling is observed[54,59]. Therefore, we also consider two CrI3 layers separated by a non-magnetic layer and simulated the antiferromagnetic interlayer coupling as a RKKY interaction added to the effective field, hence as a standard SAF $\mathbf{h}_{ex,i}^{inter} = \frac{A^{ex}}{\mu_0 M_s^{j2} t_{NM}}\mathbf{m}^j$ [60], where $i,j = L, U$, $A_{ex}$ is the interlayer exchange coupling constant, $\mu_0$ is the vacuum permeability, and $t_{NM}$ is the thickness of the non-magnetic layer. All the simulations were performed at zero bias field and temperature.

| Parameters | Nèel Skyrmion | Bloch Skyrmion | Radial Vortex | Circular Vortex |
|---|---|---|---|---|
| $M_s$ (kA/m) | 900 | 900 | 1000 | 1000 |
| $A$ (pJ/m) | 20 | 20 | 20 | 20 |
| $K_u$ (MJ/m$^3$) | 0.8 | 0.8 | 0.5 | 0.5 |
| $|D|$ (mJ/m$^2$) | [1.5, 3.0] (square & circular geometries) | [1.5, 3.0] (square & circular geometries) | [1.67, 2.00] (rectangular geometry) [1.70, 2.30] (circular dot) | [0, 0.8] (rectangular geometry) [0, 1.1] (circular dot) |

**Table 1** Micromagnetic parameters of *CoFeB* used for simulating skyrmions [20] and magnetic vortices [34].

| Parameters | Néel Skyrmion | Bloch Skyrmion | Radial Vortex | Circular Vortex |
|---|---|---|---|---|
| $M_s$ (kA/m) | 68.781 | 68.781 | 68.781 | 68.781 |
| $A$ (pJ/m) | 1.21 | 1.21 | 1.21 | 1.21 |
| $K_u$ (MJ/m$^3$) | 0.317 | 0.317 | 0** | 0** |
| $|D|$ (mJ/m$^2$) | [0.59, 0.78] (square & circular geometries) | [0.60, 0.77] (square & circular geometries) | [0.045, 0.070] (rectangular geometry) [0.040, 0.090] (circular dot) | No DMI range (rectangular & circular geometries) |

**Table 2** Micromagnetic parameters of *CrI$_3$* used in the simulations [53]. **The value of the anisotropy of *CrI$_3$* can be tuned as shown in Fig. 1(c) of Ref. [66].

*2.2 Thiele equation.* The core-translation of topological textures can be captured by a particle-like behavior which is well described in terms of the Thiele formalism[61,62]. In the Thiele formalism, we assume that the magnetization evolves adiabatically such that at all instants the full magnetization is uniquely defined by the position of the topological texture core, i.e. $\mathbf{m}(\mathbf{x},t) \equiv \mathbf{m}(\mathbf{x} | X(t), Y(t))$ where $X(t)$ and $Y(t)$ are the position of the soliton core. Hence, the time-evolution of the magnetization is described as $\mathbf{m}(\mathbf{x},t) \equiv \mathbf{m}(\mathbf{x} - \mathbf{V}t)$, where $\mathbf{V}(t) = \dot{X}(t)\mathbf{x} + \dot{Y}(t)\mathbf{y}$, for both skyrmions and vortices. Within these assumptions, we can integrate the LLG Eq. (1) and obtain,

$$\left(\overleftrightarrow{G} + \alpha_G \overleftrightarrow{D}\right)\mathbf{v} + \mathbf{F} = 0, \tag{6a}$$

where,

$$\left(\overleftrightarrow{G}\right)_{ij} = \varepsilon_{zij} \int \mathbf{m} \cdot \left(\partial_x \mathbf{m} \times \partial_y \mathbf{m}\right) d^2x = 4\pi Q \varepsilon_{zij}, \tag{6b}$$

$$\left(\overleftrightarrow{D}\right)_{ij} = \int \left(\partial_i \mathbf{m} \cdot \partial_j \mathbf{m}\right) d^2 x , \quad (6c)$$

$$\mathbf{F} = -\frac{1}{\mu_0 M_s^2 t} \nabla E . \quad (6d)$$

Here, Q is the topological charge and is independent of the precise shape of the magnetization configuration. $Q = \pm 1$ for skyrmions and $Q = \pm 0.5$ for vortices according to the soliton polarity. Moreover, $\varepsilon_{ijk}$ is the antisymmetric tensor with $\varepsilon_{xyz} = 1$, and $E$ is the total free energy. $\overleftrightarrow{G}$ and $\overleftrightarrow{D}$ are called gyrotropic and viscosity tensors, respectively.

The gyrotropic tensor $\overleftrightarrow{G}$ produces a motion perpendicular to the applied force and depends only on the topological charge of the soliton [20,44,62–64]. In the case of the RKKY coupled system, the two solitons have opposite topological charge, and thus, the collective soliton experiences the motion corresponding to zero topological charge[60,65]. Thus, the net velocity of the collective solitons have no component perpendicular to the force. The viscosity tensor $\overleftrightarrow{D}$ produces a motion along the direction of the force, leading to a minimization of the total free energy E. Eqs. 6(c)-(d) depend on the exact ansatz of the soliton with the major contributions coming from the vicinity around the soliton core, where the magnetization is not in plane and has a non-vanishing gradient. The spatial dependence of the total free energy $E$ (see Eq. 6(d)) may be due to edge repulsion and material inhomogeneities.

*2.3 Sample properties.* The material parameters used in simulations are detailed in Table I for CoFeB and in Table II for CrI3. The different anisotropies for the skyrmion and vortices simulations are essential for the stabilization of the respective topological solitons, and are allowed due to the tunability of the anisotropy with proper sample design [66].

For the skyrmion-hosting samples, we analyze a circular geometry with diameter from $d = 100$ nm to 500 nm as well as a 100 nm x 100 nm square. The thickness was set to $t_{FM} = 0.31$ nm. The discretization cell used in both geometries is 1 nm x 1 nm x 0.31 nm.

For vortices-hosting samples, we again consider circular geometry with diameter from $d = 100$ nm to 500 nm as well as a strip with dimensions 1500 nm x 250 nm. The thickness was set to $t_{FM} = 1$ nm. The discretization cell used in both geometries is 5 nm x 5 nm x 1 nm. The same results are obtained with a thickness of 0.31 nm and a discretization cell 1 nm x 1 nm x 0.31 nm. We also simulated a SAF for each of the two materials (CoFeB, CrI3), composed of two ferromagnets separated by a non-magnetic layer with thickness 2 nm for CoFeB and 0.62 nm for CrI3.

## 3. Stability Results

In this section, we perform static simulations to obtain the DMI range for the stability of the different solitons. This range of DMI will be employed as a linear gradient for the soliton manipulation.

*3.1 Skyrmions.* Magnetic skyrmions stabilization is well-known in thin-film and occurs in the absence of external magnetic fields by the interplay of exchange, DMI, uniaxial anisotropy and magnetostatic fields. It has been shown that isolated skyrmions are metastable states for a range of DMI given by $\sim 0.5\ D_c < D < D_c$ where $D_c = 4\sqrt{A(K_u - 0.5\mu_0 M_s^2)}/\pi$ [58]. In the thin film approximation, this range is mostly independent of the DMI type, bulk or interfacial. For the *CoFeB* values in Table I, we obtain $D_c = 3$ mJ/m$^2$, while for *CrI$_3$* we obtain $D_c = 0.78$ mJ/m$^2$. These theoretical predictions were corroborated by the static micromagnetic investigation as shown in the stability range in Table I and II. These ranges are independent of the sample geometry. The reason is that skyrmions are localized textures and are not affected by edge effects when they are much smaller than the sample and far from the edges.

*3.2 Vortices.* Magnetic vortices are non-local magnetic textures and, thus, are highly influenced by the sample geometry, boundary conditions, and their chirality (radial or circular). We observed in the micromagnetic simulations that radial vortices are stabilized in both materials within a DMI range which depends on the geometry. In particular, the radial vortex is stabilized for circular dot diameters larger than 250 nm, as reported in the literature [34]. At a 500 nm diameter, the radial vortex is stable for 1.5 mJ/m$^2 \leq |D| \leq 2.1$ mJ/m$^2$ (not shown). However, we notice that circular vortices are stabilized only in the *CoFeB* thin film (0.0 mJ/m$^2 < |D| < 1.1$ mJ/m$^2$), within a DMI range in agreement with the literature [34]. Figure 1 illustrates the stability of radial vortices as a function of *D* in the 1500 nm x 250 nm rectangular sample of *CoFeB* (Figs. 1(a)-(f)) and *CrI$_3$* (Figs. 1(g)-(n)). For the *CoFeB*, we notice that in the range 0.0 mJ/m$^2 \leq |D| \leq 0.8$ mJ/m$^2$, circular vortices are stabilized, and, in the range 1.67 mJ/m$^2 \leq |D| \leq 2.0$ mJ/m$^2$, radial vortices are stabilized. For the *CrI$_3$*, radial vortices are stabilized in the range 0.049 mJ/m$^2 \leq |D| \leq 0.070$ mJ/m$^2$.

We observe that the interplay of exchange, anisotropy and i-DMI play an important role in the stability of vortices, that changes according to the considered geometry (circular or rectangular). For the upper boundary of radial vortices, we find the transition to the helical state, see Fig. 1(f) and (n). Moreover, we notice that the range of i-DMI for the radial vortex stability in *CrI₃* is around 30 times smaller than in the *CoFeB*.

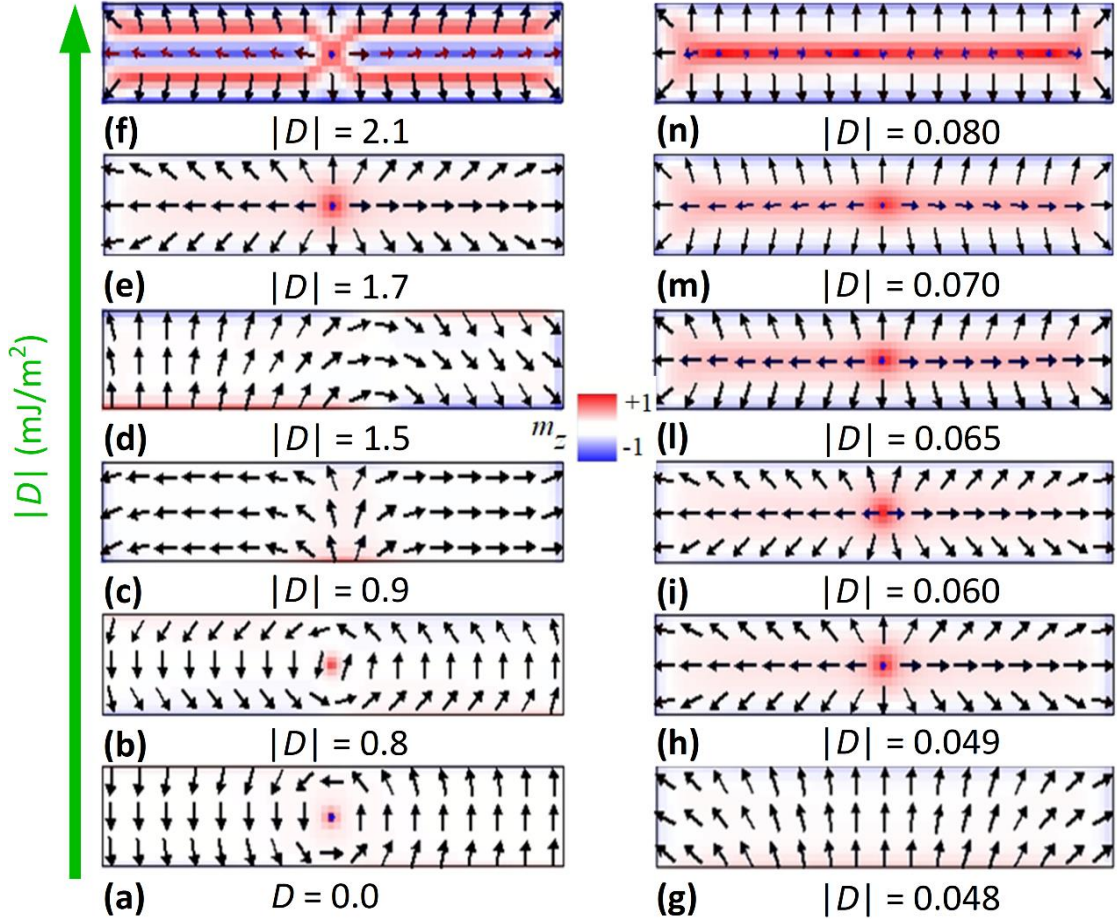

**Figure 1.** Micromagnetic simulations results of the equilibrium configurations of the magnetization as a function of |D| in (a) – (f) *CoFeB* and (g)-(n) *CrI₃* rectangular strips.

4. **Dynamics results**

In this section, we performed micromagnetic simulations considering gradients of DMI within the range of stability of the magnetic solitons in the respective samples, as obtained in Section 3. In particular, we fix the minimal and maximal values of the DMI as the values at the edge of the sample and consider a linear gradient along the x-direction of the sample.

*4.1 Skyrmions.* As mentioned in the introduction, the motion of Néel skyrmions in DMI gradients have been already analyzed in the literature with parameters similar to the ones considered here for *CoFeB* [44,45]. For this reason, we focus on the 2d material *CrI₃* and analyze both Bloch and Néel skyrmion dynamics induced by the linear gradient of b-DMI and i-DMI, respectively (see Figs. 2(a)-(b)).

We start from a single layer *CrI$_3$*, which reveals a ferromagnetic behavior[54,59]. For the chosen $K_u$ and $M_s$, the Néel and Bloch skyrmions move with positive velocities in the x- and y-directions. Along the x-direction, it moves towards a higher DMI value, while the motion along the y-direction is given by the Magnus force[44]. This is in agreement with the Thiele equations for skyrmions[41,44]. The damping coefficient $\alpha_G$ is responsible for the motion towards a lower energy state, corresponding to a higher DMI. Hence, in the simulations, we notice that a higher damping constant is associated to a higher component of the velocity along the linear gradient.

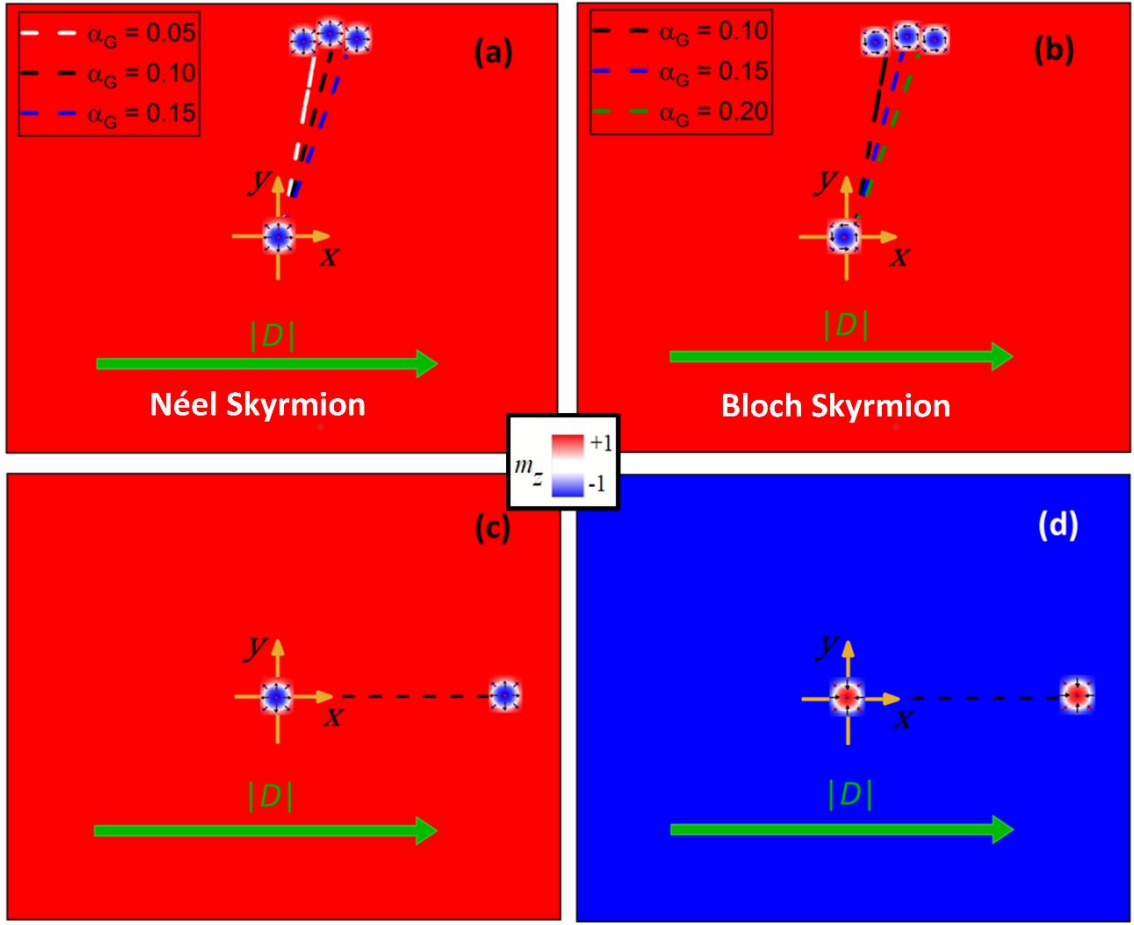

**Figure 2.** Micromagnetic simulations results of the (a) Néel skyrmion trajectory under a linear i-DMI gradient (0.59 mJ/m2 ≤ D ≤ 0.78 mJ/m2, see Table I), and (b) Bloch skyrmion trajectory under a linear b-DMI gradient (0.60 mJ/m2≤D≤ 0.77 mJ/m2 directed along the positive x-direction as shown in green arrow in a single CrI3 layer as a function of different damping coefficients. (c) – (d) Top and bottom layer of the SAF CrI3, respectively, where the straight trajectory of the skyrmion is also indicated. The color bar denotes the out-of-plane magnetization component mz: red positive, white zero, and blue negative.

We proceed to consider a *CrI$_3$* with two layers, which coupled antiferromagnetically[54,59] and can be modeled as a SAF[60]. In the SAF *CrI$_3$*, two skyrmions of opposite topological charges are antiferromagnetically coupled via the interlayer exchange coupling. This coupling leads to a zero net Magnus force and a suppression of the skyrmion Hall effect [44,60] which results in a straight motion along the gradient direction (Fig. 2(c)-(d)). The observed behaviors are in agreement with the Thiele formalism developed in Sec. 2.2 by employing the skyrmion ansatz as described in Ref. [44]. We conclude that the skyrmion Hall angle in 2d *CrI₃* depends on the number of layers, and it is present for an odd number of layers but vanishes for an even number of layers.

*4.2 Radial Vortex*. Figures 3(a)-(b) show the dynamics induced by the i-DMI gradient on a single layer 1500 nm x 250 nm *CoFeB* and *CrI₃* rectangular strips. In both materials, the vortex core exhibits a translational motion under the influence of the i-DMI gradient. We notice that it moves towards the region of higher $|D|$, with a significant Hall angle. Similarly to skyrmions, we notice that velocity depends on the damping coefficient, with smaller y-component for higher damping values. However, we observe a major difference: while a skyrmion moves mainly in the positive y-direction, a radial vortex tends to move mainly in the positive x-direction. This is ascribed to: (i) the topological charge of vortices is half of the skyrmion, and (ii) vortices are more influenced by sample boundary effects than skyrmions, since they are non-local textures.

In a SAF (either *CoFeB* or *CrI₃* with an even number of layers), the radial vortex moves with a zero skyrmion Hall angle (Fig. 3(c)-(f)). However, the radial vortex travels faster in the case of *CrI₃* (compare Figs. 3(c) and (e) for *CoFeB* with Figs. 3(d) and (f) for *CrI₃* obtained within the same 250ns time interval). The same behavior is observed in a circular sample.

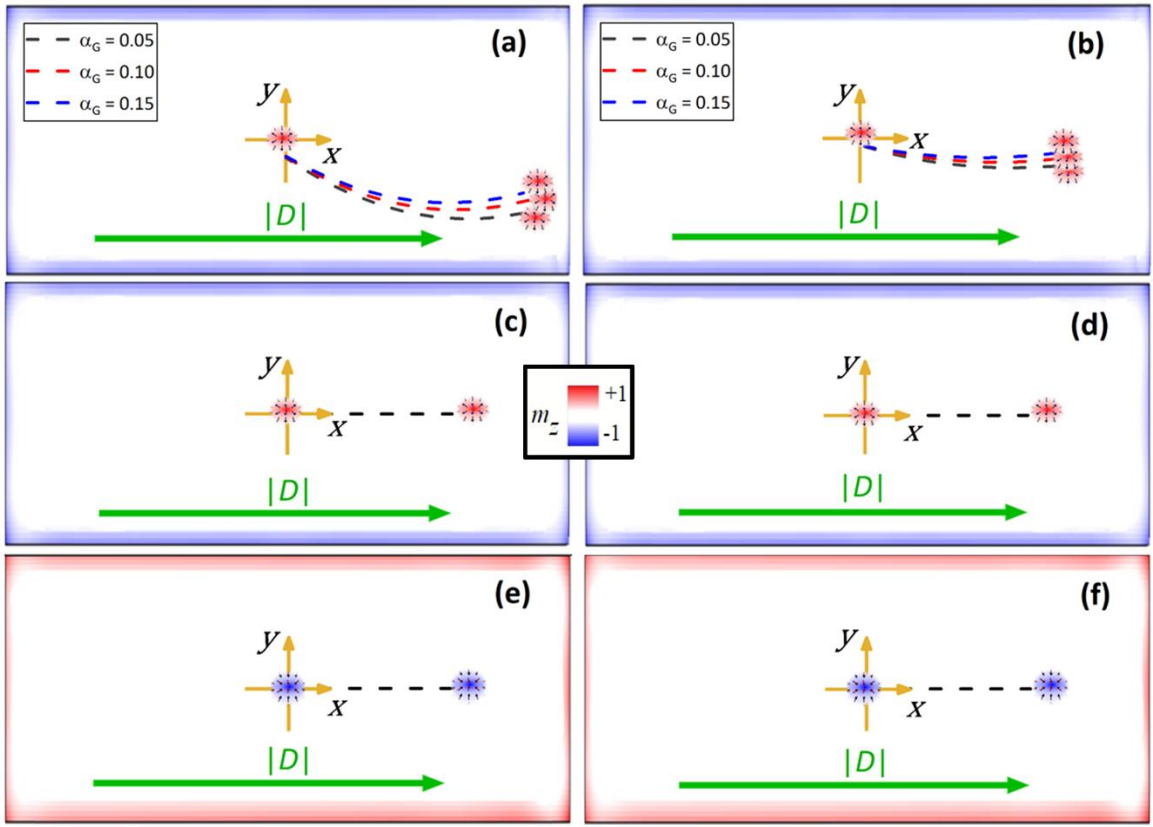

**Figure 3.** Micromagnetic simulations results of radial vortex trajectories under a linear i-DMI gradient. In (a) and (b), a single 1500 nm x 250 nm FM rectangular sample under the effect of different damping coefficients in *CoFeB* and *CrI$_3$*, respectively, (c) and (d) in top layers and (e) and (f) in the bottom layers of the SAF composed of *CoFeB* and *CrI$_3$*, respectively.

For a precise quantitative analysis based on the Thiele equation (see Eqs. 6(a)-(d)), it is necessary to obtain the full magnetic configuration of the magnetic vortex. Previous studies have considered numerical approaches to compare to the micromagnetic simulations[62,63]. Here, we consider a qualitative discussion and, for simplicity, we assume that, in the region with major contribution to the integrals in Eqs. 6(c)-(d), the magnetization configuration is given by a radially symmetric configuration. The radial symmetry implies $(\vec{D})_{xx} \approx (\vec{D})_{yy} \approx \Gamma$, $(\vec{D})_{xy} \approx (\vec{D})_{yx} \approx 0$.

A qualitative description of the vortex motion is shown in Fig. 4, which is in agreement with the simulation results of Fig. 3(a). We considered both variations of the damping and the DMI gradient. Moreover, up to the lowest non-zero order, we can write the free energy of the vortex, due to the DMI gradient and the border repulsion, as $E = -c_D g_D X + kY^2$, where $X, Y$ are the coordinates of the center of the vortex along the x and y directions. We also have that $c_D \approx \int d^2x \left[ m_z \nabla \cdot \mathbf{m} - (\mathbf{m} \cdot \nabla) m_z \right]$ and $k \approx \frac{1}{2} \partial_Y^2 \int d^2x \left( \mathbf{m} \cdot \mathbf{h}_{bound}(x,y,Y) \right)$ are constants, assuming that the magnetization $\mathbf{m}(x)$ is invariant of the position (we do not consider strong deformations of the texture as it moves) and $\mathbf{h}_{bound}(x,y,Y)$ represents the effective magnetic field due to boundary effects. Moreover, we assumed $D(x) = g_D x$ as the linear gradient of the i-DMI parameter. The Thiele equations of motion are,

$$\dot{X} = \frac{1}{\mu_0 M_s^2 t} \frac{-2\pi kY + \alpha_G c_D g_D \Gamma}{4\pi^2 + \alpha_G^2 \Gamma^2}, \quad (7a)$$

$$\dot{Y} = -\frac{1}{\mu_0 M_s^2 t} \frac{2\pi c_D g_D + \alpha_G k \Gamma Y}{4\pi^2 + \alpha_G^2 \Gamma^2}. \quad (7b)$$

We observe that, due to the DMI gradient, the vortex is pushed towards the bottom edge, while the dissipative term pushes it along the gradient. Additionally, the edge repulsion tends to accelerate the skyrmion along the x-direction as it approaches the edge while forcing the vortex back to the center of the strip due to a small dissipative contribution.

As a remark, we notice that lower anisotropy tends to lower the velocity component along x and increase the component along y, while a lower DMI gradient tends to decrease the velocity along x and along y.

*4.3 Circular Vortex.* We perform a similar analysis to observe the effect of the i-DMI gradient on circular vortices, remarking that the circular vortex has not been found stable for the parameters of the 2d *CrI$_3$* (Table II).

According to our observations, the circular vortex undergoes an expulsion dynamics under the influence of the linear i-DMI gradient (Fig. 5). Since the circular vortex is not a stable magnetization configuration under any finite value of *D*, due to the damping, it decays to the uniform in-plane state [44]. This mechanism in interesting for application of spintronic devices as diodes[67,68]. Recently, it has been shown that the vortex core expulsion can be a mechanism driving an enhancement of sensitivity up to 80kV/W [42, 49].

Thus, we envision that the recent discovery of voltage controlled DMI can drive additional enhancement of the performance of spintronic diodes by exploiting this behavior.

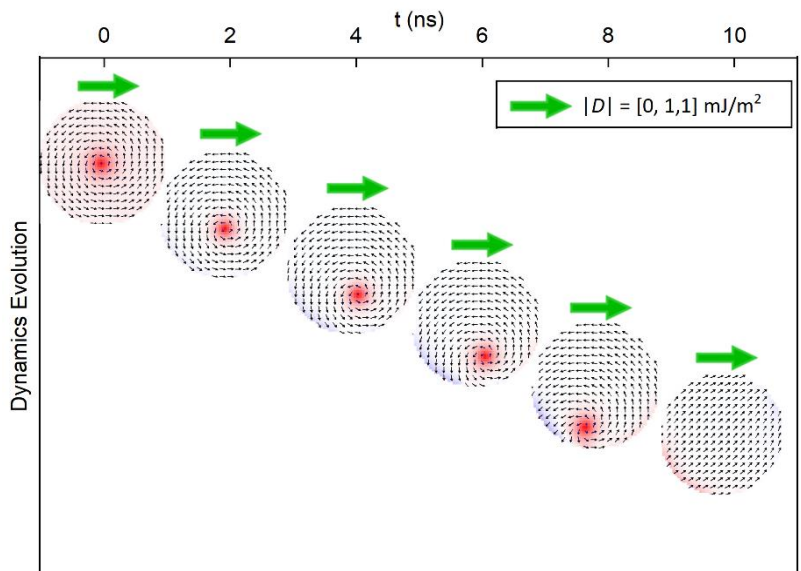

**Figure 5.** Micromagnetic simulations results of the time evolution of the spatial distribution of the magnetization for a circular vortex under the influence of the linear i-DMI gradient in a *CoFeB* circular sample

## 5. Conclusion

In summary, we performed a systematic study of the effect of linear DMI gradient on skyrmions (Bloch and Néel types), radial and circular vortices motion in a *CoFeB* thin-film and a 2d *CrI$_3$* through micromagnetic simulations and Thiele's equation. Our results showed that stabilization of these magnetic configurations can take place for specific intervals of DMI (depending on the sample geometry) at zero external field. Under the effect of a linear i-DMI gradient, skyrmion and radial vortex move with a damping-dependent trajectory in a single-layer *CoFeB* and 2d *CrI$_3$*, while a zero skyrmion Hall angle motion was observed in the SAF *CoFeB* and *CrI$_3$* for both skyrmions and vortices.

On the contrary, the circular vortex was expelled from the sample since it is not stable for any finite i-DMI, and we suggested that these dynamics can be exploited in voltage controlled DMI spintronic diodes to increase performances. Our results suggest alternative means for low-power manipulation of magnetic solitons in FM and SAF, and, particularly, in 2d materials. Specifically, future developments on the control of magnetic parameters by strain, geometry design, temperature gradients, and applied voltages and the implementation of 2d materials combined with soliton manipulation can allow for a new generation of highly efficient sensors and diodes for computing and energy harvesting. 0

**Acknowledgements**

This work was supported under the project number 101070287 — SWAN-on-chip — HORIZON-CL4-2021-DIGITAL-EMERGING-01, the project PRIN 2020LWPKH7 funded by the Italian Ministry of University and Research, and by PETASPIN association (www.petaspin.com).

**Keywords**

skyrmions, vortices, 2d materials, gradients